\documentclass[aps,prd,12pt,nofootinbib]{revtex4-2}

\usepackage[bookmarksnumbered, pdfpagelabels=true, plainpages=false, colorlinks=true, linkcolor=blue, citecolor=blue, urlcolor=blue]{hyperref}

\usepackage{graphicx}
\usepackage{animate}
\usepackage{bm}

\usepackage{subfloat}
\usepackage{float}
\usepackage{amssymb}
\usepackage{amsmath}
\usepackage[usenames]{color}

\newcommand{\be}{\begin{equation}}
\newcommand{\ee}{\end{equation}}
\newcommand{\ba}{\begin{eqnarray}}
\newcommand{\ea}{\end{eqnarray}}
\newcommand{\bd}{\begin{displaymath}}
\newcommand{\ed}{\end{displaymath}}

\def\thalf{{\textstyle{\frac{1}{2}}}}

\def\oneqt{{\textstyle{\frac{1}{4}}}}

\begin{document}


\title{Relaxation times for disoriented isospin condensates in high energy heavy ion collisions}

\author{Olivia Chabowski}
\email{chabo026@umn.edu}
\affiliation{School of Physics \& Astronomy, University of Minnesota, Minneapolis, MN 55455, USA}

\author{Joseph I. Kapusta}
\email{kapusta@umn.edu}
\affiliation{School of Physics \& Astronomy, University of Minnesota, Minneapolis, MN 55455, USA}

\author{Mayank Singh}
\email{mayank.singh@vanderbilt.edu}
\affiliation{School of Physics and Astronomy, University of Minnesota, Minneapolis, MN 55455, USA}
\affiliation{Department of Physics and Astronomy, Vanderbilt University, Nashville, TN 37240, USA}

\begin{abstract}
Fluctuations between charged and neutral kaons measured by the ALICE Collaboration in Pb-Pb collisions at the LHC exceed conventional explanations. Previously it was shown that if the scalar condensate is accompanied by an electrically neutral isospin--1 field then the combination can produce large equilibrium fluctuations where $\langle \bar{u}u\rangle \ne \langle \bar{d}d\rangle$. Hadronizing strange and anti-strange quarks might then strongly fluctuate between charged ($u\bar{s}$ or $s\bar{u}$) and neutral ($d\bar{s}$ or $s\bar{d}$) kaons.  Here we estimate the times for the condensates to achieve their equilibrium probability distributions within causal volumes in high energy heavy ion collisions.  This is achieved by modeling the temperature dependence of the condensates, mesonic collective excitations, decay rates of the associated fields, and employing the Langevin and Fokker-Planck equations.  Within this model, we find that the equilibration times are short compared with the expansion time, making disoriented isospin condensates a viable explanation for the anomalous fluctuations observed at the LHC. 
\end{abstract}

\maketitle

\section{Introduction}\label{SecI}

The ultrarelativistic heavy-ion collision experiments are a class of particle collider experiments that have been in operation since 2008 \cite{PHOBOS:2000wxz,STAR:2000ekf,PHENIX:2000owy,BRAHMS:2001gci,Hallman:2002qi}. This class includes experiments at the Large Hadron Collider (LHC) involving collisions between lead ions (Pb-Pb collisions), which have been running since 2010 \cite{ALICE:2010mlf,CMS:2011aqh,ATLAS:2011ag,Schutz:2011zz}. Heavy-ion collision experiments aim to create nuclear matter at extreme temperatures and energies that mimic the universe's conditions about a microsecond after the Big Bang. These experiments explore the fundamental properties of nuclear matter in extreme conditions.

A salient feature of Quantum Chromodynamics (QCD), the theory of strong nuclear force, is confinement. At the low temperatures found in the contemporary universe, quarks and gluons are only observed in bound states called hadrons. At high temperatures these quarks and gluons can be deconfined from hadronic states to form quark-gluon plasma (QGP). Lattice QCD calculations show that this change in phase is a smooth crossover at low baryon densities \cite{Aoki:2006we}, which is relevant for LHC collision energies.  The QGP formed in these collisions expands and cools and can be described as a relativistic dissipative fluid \cite{Heinz:2013th,Gale:2013da}. As the QGP reaches the crossover temperature, the partons again go into confinement forming a gas of hadrons.

In the quark confinement phase of QCD, the vacuum expectation value of a quark and its antiparticle is non-zero and is referred to as a quark condensate. The up and down quark condensates are expected to be equal-valued in this phase. These quark condensates ``melt’’ at high temperatures, meaning that their values go to zero \cite{Lee:1993aw}. This corresponds to the approximate chiral symmetry restoration of QCD. As the system temperatures fall in the aftermath of a Pb-Pb collision, the condensates re-form and the approximate chiral symmetry is broken again. As the condensates re-form, it is plausible that they temporarily fluctuate from their equilibrium values. As the hadronization happens in the same temperature range as the chiral symmetry restoration, these fluctuations can leave their signatures on the hadron distributions. Isolating such signatures would lead to experimental evidence of approximate chiral symmetry restoration at high temperatures in QCD.

A possible condensate fluctuation could come if the up and down condensates are not equal-valued. Small domains could form within the cooling QGP where $\langle u \bar{u}\rangle \neq \langle d \bar{d}\rangle$. Such a fluctuation is referred to as a disoriented isospin condensate (DIC) and was proposed in \cite{Kapusta:DIC}. It was postulated that the DICs would be possible if the iso-scalar field, $\left( \langle u \bar{u} \rangle + \langle d \bar{d}\rangle \right)/\sqrt{2}$, were accompanied by the iso-vector field, which has components $\langle u \bar{d} \rangle$, $\langle d \bar{u} \rangle$, and $ \left( \langle u \bar{u} \rangle - \langle d \bar{d}\rangle \right)/\sqrt{2}$. The lowest excitation of the iso-scalar field is associated with the $f_{0} (500)$, or the $\sigma$ meson while the lowest excitations of the iso-vector field are associated with the three variants of the $a_{0} (980)$ meson, $a_{0}^{+}$, $a_{0}^{-}$, and $a_{0}^{0}$ mesons. While the vacuum condensate is an iso-scalar singlet, it can easily rotate in an iso-vector triplet direction if the free energy barrier is small. It was shown within the linear sigma model that the energy cost of disorientations of the iso-vector field is high at vacuum but is reduced significantly above the crossover temperature $T_{h}$. The hadronization of strange quarks in $\langle u \bar{u}\rangle$ rich domains would lead to a relative excess of charged kaons as compared to neutrals kaons while their hadronization in $\langle d \bar{d}\rangle$ rich domains would yield the opposite result.

The DIC was proposed to explain the kaons flavor fluctuations measured by the ALICE collaboration  \cite{ALICE:2021fpb}. The dynamical correlator $\nu_{\textnormal{dyn}}$  \cite{Gavin:2001uk} was used to measure correlations between charged and neutral kaons on an event-by-event basis in Pb-Pb collisions at $\sqrt{\textnormal{s}_{\textnormal{NN}}} = 2.76$ TeV. The $\nu_{\textnormal{dyn}} \left[A,B\right]$ has the following form:
\begin{gather}
    \nonumber \nu_{\textnormal{dyn}} \left[A,B\right] = R_{AA} + R_{BB} - 2 R_{AB} \\
    R_{xy} = \frac{\langle N_{x} \left( N_{y} - \delta_{xy} \right) \rangle}{\langle N_{x} \rangle \langle N_{y} \rangle}
\end{gather}
Here $N_{A}$ refers to the number of observed particles of type A, and $\langle ... \rangle$ refers to averages over events. While the correlations between kaons of opposite charges exhibited behavior largely consistent with the predictions of standard heavy-ion collision simulators, the correlations between charged and neutral kaons deviated significantly from predictions. Values of $\nu_{\textnormal{dyn}}[\textnormal{K}_{s}^{0},\textnormal{K}^{\pm}]$ were measured to be significantly larger than the values predicted by HIJING, AMPT and EPOS–LHC \cite{Nayak:2019qzd}. 
These anomalously large correlations were not local and persisted over a unit in rapidity. The ALICE collaboration also scaled $\nu_{\rm dyn}$ by $\alpha \left[A,B\right] \equiv \langle N_{A} \rangle^{-1} + \langle N_{B} \rangle^{-1} $, which is proportional to inverse multiplicity and roughly scales with the number of sources of two-particle correlations. 
Multiplying $\nu_{\textnormal{dyn}}$ by $1/\alpha$, they found that $\left( \nu_{\textnormal{dyn}}/\alpha \right) \left[ \textnormal{K}_{s}^{0},\textnormal{K}^{\pm} \right]$ was higher for central collisions, while HIJING, AMPT and EPOS–LHC all predicted $\left( \nu_{\textnormal{dyn}}/\alpha \right) \left[ \textnormal{K}_{s}^{0},\textnormal{K}^{\pm} \right]$ to be independent of centrality.

It was shown that the conventional mechanisms like resonance decays, charge conservation, and Bose symmetrization were all insufficient to explain the anomalously large magnitude, centrality dependence and long-range nature of $\left( \nu_{\textnormal{dyn}}/\alpha \right) \left[ \textnormal{K}_{s}^{0},\textnormal{K}^{\pm} \right]$\cite{Kapusta:2022ovq}. The measurements could be explained if a significant number of kaons were produced in domains where the neutral fraction of kaons was equally likely to take any value between zero and one.

If DICs coupled with strange quark hadronization is the process leading to domains of primarily neutral or primarily charged kaons, the causally connected volumes that can form a single domain must be accounted for. Correlated kaon emissions cannot originate from causally disconnected regions of space, a limitation described by Castorina and Satz as ``the counterpart of the cosmological horizon problem'' \cite{CS}. The cosmological horizon problem arises from the observation of extremely similar cosmic microwave background temperatures in regions of space thought to be causally disconnected, and it is one of the problems addressed by cosmological inflation theory. It is therefore necessary to model the temporal evolution of causal volumes in Pb-Pb collisions in describing resultant kaon fluctuations.

In this work, we calculate the formation times of DICs in a causally connected domain and show that equilibrated DICs can form on a very short time scale compared to the lifetime of heavy-ion collisions. This is done by treating the isospin fluctuation amplitude $\varphi$ as a field and evolving it via the Langevin and Fokker-Planck equations with the simplification that the fields are spatially uniform. The simplification helps solve the equation semi-analytically. The condensate fluctuations are related to scalar mesonic fields, which are taken to be excitations of vacuum. Our results show that the DICs are a plausible explanation of the anomalous kaon correlations and if confirmed by further experiments have the potential to provide the elusive experimental evidence of approximate chiral symmetry restoration of QCD.

In section \ref{SecII}, we describe our temperature-dependent parameterization of quark masses, meson masses, and the scalar field condensates $\bar{\sigma}$ and $\bar{\zeta}$, which we obtain by coupling quarks to the iso-scalar and iso-vector fields, as well as by making use of lattice QCD computations of the temperature dependence of $\langle u \bar{u} \rangle$ and $\langle d \bar{d} \rangle$. We model the decay rates of the $\sigma$ and $\zeta$ fields by interpreting mesons as finite-temperature collective excitations of the vacuum in section \ref{SecIII}. In section \ref{SecIV}, we analyze how the potential of a spatially uniform meson field depends on fluctuations in the balance between $\sigma_{u}$ and $\sigma_{d}$, to which we assign the parameter $\varphi$. We demonstrate that, in the strong damping limit, our system behaves in accordance with the well-understood Langevin equation and obtain the corresponding Fokker-Planck equation in section \ref{SecV}. In section \ref{SecVI}, we analyze how the Fokker-Planck equation describing our system changes if we do not suppress the second-order time derivative of fluctuations in $\varphi$, as we did when considering the strong damping limit. In section \ref{SecVII}, we apply a method devised by Chandrasekhar \cite{Chandrasekhar} to the Fokker-Planck equation we obtained in the previous section, assuming a spatially uniform region where  $\varphi$ varies in time. In section \ref{SecVIII}, we parameterize the evolution of temperature using Pb-Pb collision simulation data, taking into account the time-dependent dimensions of expansion. We model the time and temperature dependence of the size of Pb-Pb collision causal volumes and the propagation speed of excitations in $\varphi$ in section \ref{SecIX}. The results are summarized and discussed in section \ref{SecX}.

\section{Condensates and Masses}\label{SecII}

In the temperature range of interest, which is approximately $160 < T < 400$ MeV, the matter is strongly interacting, and perturbation theory cannot reliably be applied.  We address this problem by coupling quarks to the scalar fields $\sigma$ and $\zeta$, such as in the NJL (Nambu--Jona-Lasinio) model \cite{NJL1,NJL2}.  Condensation of these fields gives rise to constituent quarks.  In a flavor basis the scalar field is $\Sigma = {\rm diag} (\sigma, \sigma, \sqrt{2} \zeta)$.  The current quark mass matrix is ${\cal M} = {\rm diag} (m_u, m_d, m_s)$.  Then
\be
{\cal L}_q = \bar{q} \left( i \! \not\!\partial - {\cal M} - g \Sigma \right) q
\ee
Denoting the condensates by $\bar{\sigma}$ and $\bar{\zeta}$, the constituent quark masses are
\ba
M_u &=& m_u + g \bar{\sigma} \nonumber \\
M_d &=& m_d + g \bar{\sigma} \nonumber \\
M_s &=& m_s + \sqrt{2} g \bar{\zeta}
\ea
The zero temperature values of the condensates are $\bar{\sigma} = f_\pi$ and $\sqrt{2} \bar{\zeta} = 2 f_K - f_\pi$.  From now on we take $m_u =m_d \equiv m_l$.  According to the PDG (Particle Data Group) \cite{PDG}, $f_\pi = 92.1$ MeV and $f_K = 110.1$ MeV.

The temperature dependence of the quark condensates has been computed with lattice QCD \cite{HotQCD1}.  
The QCD chiral phase transition is in the same universality class as the 3D Ising model. In the mean-field approximation, the magnetization $m$ in the Ising model is given by
\be
m = \tanh\left(\frac{mT_c+h}{T}\right)
\ee
where $T_c$ is the critical temperature and $h$ is the external field. We can fit the condensate values from the lattice calculations using this functional form.  We parameterize the relative size of the light quark condensate $\bar{\sigma}(T)/\bar{\sigma}(0)$ as
\be
\frac{\bar{\sigma}(T)}{\bar{\sigma}(0)} = \tanh\left[\frac{1}{T}\left(\frac{\bar{\sigma}(T)}{\bar{\sigma}(0)}T_{cl}+h_l\right)\right] 
\ee

\begin{figure}[h]
\includegraphics[width=0.7\textwidth]{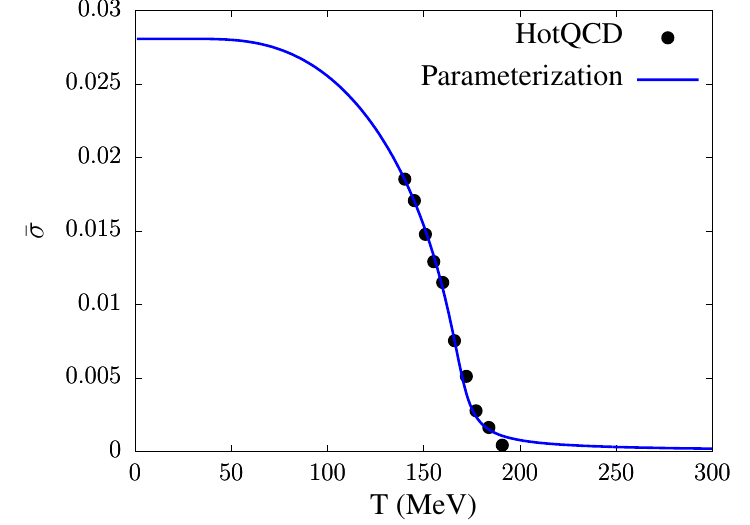}
\caption{Fit to light quark condensate with $\bar{\sigma}(0) = 0.0280795$ in lattice units}
\label{fig:ising_l}
\end{figure}

\begin{figure}[h]
\includegraphics[width=0.7\textwidth]{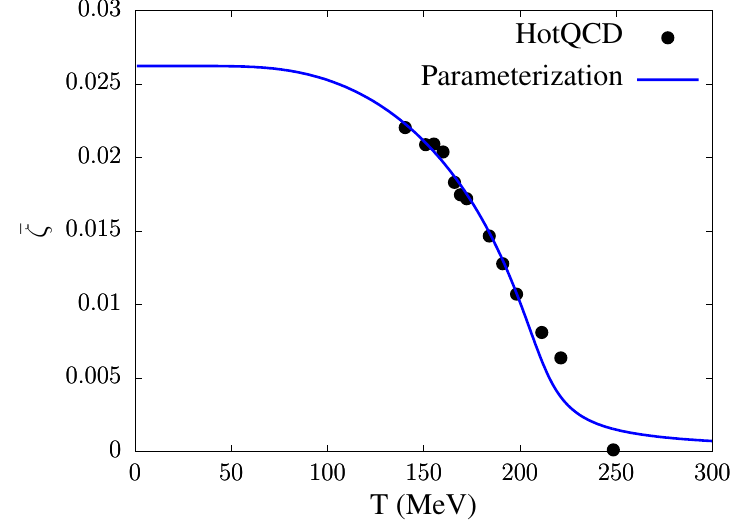}
\caption{Fit to strange quark condensate with $\bar{\zeta}(0) = 0.0262297$ in lattice units}
\label{fig:ising_s}
\end{figure}

Here $\bar{\sigma}(0) = f_\pi$. A similar parameterization for the strange quark condensate will yield 
\be
\frac{\bar{\zeta}(T)}{\bar{\zeta}(0)} = \tanh\left[\frac{1}{T}\left(\frac{\bar{\zeta}(T)}{\bar{\zeta}(0)}T_{cs}+h_s\right)\right]
\ee
with $\bar{\zeta}(0) = (2f_K - f_\pi)/\sqrt{2}$.
We obtain the best fit values $T_{cl} = 166.762$ MeV, $h_l = 0.921635$ MeV, $T_{cs} = 204.316$ MeV and $h_s = 2.6074$ MeV. The fits are shown in figures \ref{fig:ising_l}
and \ref{fig:ising_s}. The condensate values at the highest temperatures are not very accurate due to the delicate subtraction necessary to obtain them, and they are not included in our fits.  The asymptotic value for the light quark condensate is 
$\bar{\sigma}(T)/\bar{\sigma}(0) = h_l/T$.  A better approximation is $\bar{\sigma}(T)/\bar{\sigma}(0) = h_l/(T-T_{cl})$.  This is accurate to 2\% at $T=180$ MeV and to 0.1\% at $T=200$ MeV. 

The $\sigma$ meson is generally identified as the $f_0(500)$ by the PDG listings.  It has a mass in the range from 400 to 550 MeV and a broad width in the range from 400 to 700 MeV.  The pion has an isospin averaged mass of 138.0 MeV.  It is generally accepted that the small chiral symmetry breaking due to nonzero current quark masses is nearly restored at high temperatures.  The pion mass is almost constant with increasing temperature, begins increasing just below a crossover temperature of approximately 160 MeV, and subsequently rises linearly with temperature.  This is illustrated in a particular model in Ref. \cite{JJ}.  Meanwhile, the $\sigma$ mass decreases with increasing temperature, reaches a minimum near 160 MeV, and then rises, merging with the pion mass at higher temperatures.  Rather than choosing a specific model, we simply parameterize the masses as
\ba
m_\pi^2(T) &=& m_\pi^2 + \frac{a_\pi T^4}{T^2 + T_0^2} \nonumber \\
m_\sigma^2(T) &=& m_\sigma^2 + \frac{T^2}{T^2 + T_0^2} \left( a_\pi T^2 - m_\sigma^2 + m_\pi^2 \right)
\ea
At low temperature, these would represent particles, while at high temperatures they would represent collective excitations of the medium.  Physically, we might require that the 
$\sigma$ field can only decay into a pair of pions below some temperature $T_\pi$ and only into a pair of constituent quarks above some temperature $T_q$, with $T_q < T_\pi$.  If the light quark current masses were zero, there would be a second-order phase transition at some $T_c$ where $m_\sigma$ drops to zero.  In that case, one might expect that $T_q = T_\pi = T_c$.  The requirement that the decay channel into pions closes at $T_\pi$ is
\be
m_\sigma^2(T_\pi) = 4 m_\pi^2(T_\pi)
\ee
while the requirement that the minimum in $m_\sigma^2(T)$ occurs at $T_\pi$ is
\be
\left( \frac{d m_\sigma^2(T)}{dT} \right)_{T=T_\pi} = 0
\ee

\begin{figure}[h]
\includegraphics[width=0.7\columnwidth]{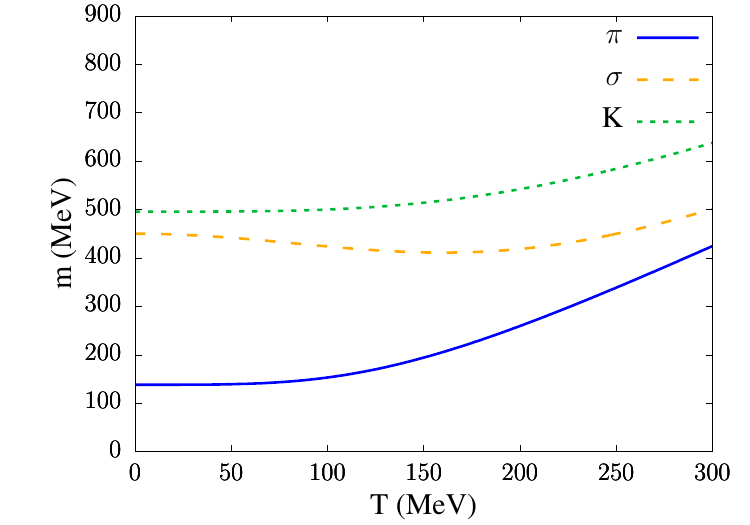}
\caption{Meson masses with $m_\sigma = 450$ MeV and $T_\pi = 160$ MeV.}
\label{fig:450_160}
\end{figure}
\begin{figure}[h]
\includegraphics[width=0.7\columnwidth]{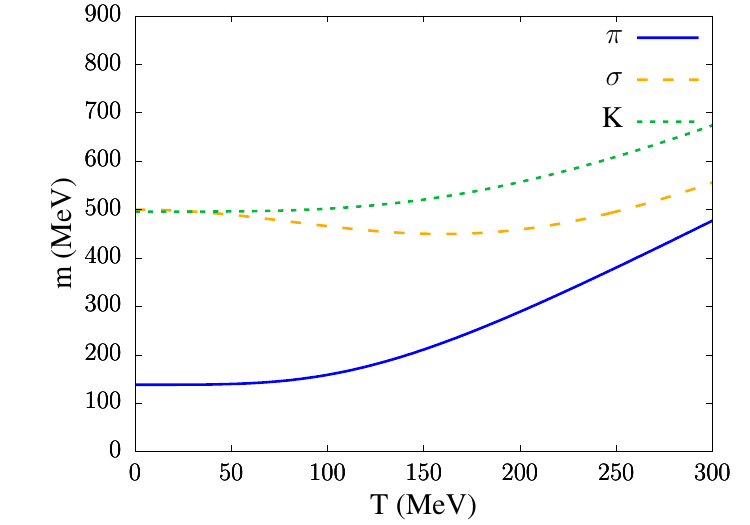}
\caption{Meson masses with $m_\sigma = 500$ MeV and $T_\pi = 160$ MeV.}
\label{fig:500_160}
\end{figure}
These determine the parameters $a$ and $T_0$ in terms of $T_\pi$. 
\be
\frac{T_0^2}{T_\pi^2} = \frac{ 2 m_\sigma^2 + 7 m_\pi^2 + 
\sqrt{ 4  m_\sigma^4 + 52 m_\pi^2 m_\sigma^2 - 47 m_\pi^4}}{4 (m_\sigma^2 - 4 m_\pi^2)}
\ee
\be
a_\pi = \frac{(m_\sigma^2 - m_\pi^2) T_0^2}{(2 T_0^2 + T_\pi^2) T_\pi^2}
\ee
The decay channel $\sigma \rightarrow \pi + \pi$ closes at $T_\pi$ by construction.  

\begin{figure}[h]
\includegraphics[width=0.7\columnwidth]{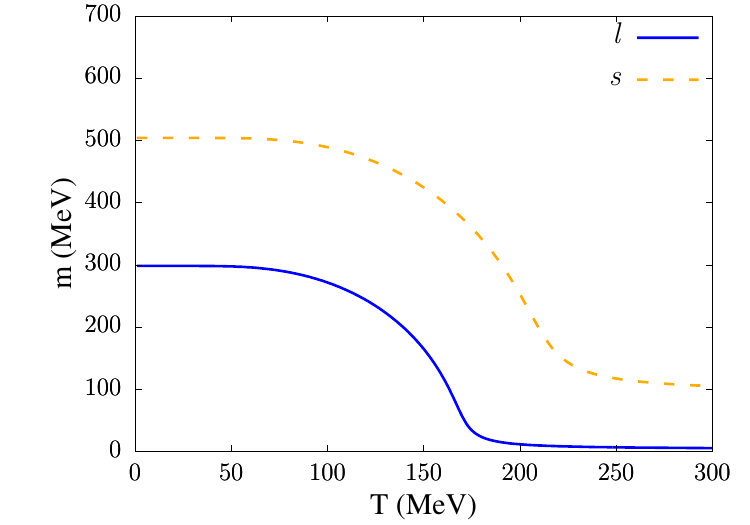}
\caption{Constituent quark masses with the parameterization as described in the text and $g = 3.2$.}
\label{fig:quark_masses}
\end{figure}

The value of $g$ is estimated by requiring that $M_l \approx 300$ MeV and $M_s \approx 500$ MeV at $T = 0$, and by ensuring that $T_q$ is close to but less than $T_\pi$, where $T_q$ is defined by
\be
m_{\sigma}(T_q) = 2 M_{l}(T_q)
\ee
This implies that $g \approx 3.2$.

The $f_0(980)$ meson is the lowest excitation of the $\zeta$ field with a T-matrix pole mass $m_{f_0}$ in the range of 980 to 1010 MeV, according to the PDG.  Like all scalar mesons, it is difficult to interpret.  Is it a quark-antiquark state, a four-quark state, a $K \bar{K}$ molecule, or some combination of these? Model calculations \cite{JJ} show that the $f_0(980)$ mass is approximately temperature independent up to $T$ = 300 MeV.  The kaon mass has a temperature dependence similar to the pion, suggesting that
\be
m_K^2(T) = m_K^2 + \frac{a_\pi T^4}{T^2 + T_0^2}
\ee
In that case, the decay mode into $K + \bar{K}$ is cut off well below $T_q$.  There is not much more we can learn from existing model studies to assist our analysis.

Figures \ref{fig:450_160} and \ref{fig:500_160} show the temperature dependence of the meson masses.
Figure \ref{fig:quark_masses} shows the temperature dependence of the constituent quark masses. 

\section{Meson Decay Rates}\label{SecIII}

Mesons may be viewed as excitations of the vacuum and interpreted and observed as particles or resonances.  Their interpretation as collective excitations at finite temperatures is certainly true as they interact with other particles/excitations in the medium.  This is clearly studied and articulated in Ref. \cite{Mocsy1}.  In this section, we estimate the decay rates of meson condensates on this understanding.  There may certainly be other processes not accounted for here, but they could only increase the rates.

The rate for a scalar particle of mass $M$ to decay into a fermion-antifermion pair of mass $m$ via a Yukawa coupling $g$ in vacuum is
\be
\Gamma_{\rm vac} = \frac{g^2 M}{8\pi} \left(1 - \frac{4m^2}{M^2} \right)^{3/2}
\ee
For the boson at rest in a finite temperature medium one must multiply by $(1 - n_F)^2$, where $n_F$ is the Fermi-Dirac occupation number, to account for Pauli blocking.  The fermions can also recombine into the boson.  Hence there is an overall factor of $(1 - n_F)^2 - n_F^2 = 1 - 2n_F$.  For the $\sigma$ field
\be
\Gamma_{\sigma qq}(T) = \frac{6g^2 m_\sigma(T)}{8\pi} \left(1 - \frac{4M_l^2(T)}{m_\sigma^2(T)} \right)^{3/2}
\left[ 1 - 2 n_F\left(\frac{m_\sigma(T)}{2T} \right) \right]
\ee
The factor of 6 comes from 2 flavors and 3 colors.  Similarly
\be
\Gamma_{\zeta ss}(T) = \frac{3g^2 m_{f_0}}{8\pi} \left(1 - \frac{4M_s^2(T)}{m_{f_0}(T)^2} \right)^{3/2}
\left[ 1 - 2 n_F\left(\frac{m_{f_0}(T)}{2T} \right) \right]
\ee
where we have taken the $f_0(980)$ mass to be temperature independent.

The decay rate for the sigma meson is
\be
\Gamma_{\sigma\pi\pi}(T) = \frac{3}{8\pi} \frac{\lambda^2 \bar{\sigma}^2(T)}{m_\sigma(T)} 
\left(1 - \frac{4m_\pi^2(T)}{m_\sigma^2(T)} \right)^{1/2}
\left[ 1 + 2 n_B\left(\frac{m_\sigma(T)}{2T} \right) \right]
\ee
The $\sigma\pi\pi$ vertex is $\lambda \bar{\sigma}$. The factor of 3 comes from the 3 decay modes into pions.  In the symmetric phase $\bar{\sigma} = 0$ so the above relations are consistent with the symmetry.  Obviously the results are dependent on the numerical value of $m_\sigma$.  In the linear sigma model one has the relation $m_\sigma^2 = 2 \lambda f_\pi^2 + m_\pi^2$ in the vacuum.
Choosing $m_\sigma = 450$ MeV in the vacuum results in $\lambda = 10.814$ and $\Gamma_{\sigma\pi\pi} = 207.82$ MeV, whereas choosing $m_\sigma = 500$ MeV results in $\lambda = 13.614$ and $\Gamma_{\sigma\pi\pi} = 312.950$ MeV.  Let's compare the two options.

\begin{figure}[t]
\includegraphics[width=0.7\columnwidth]{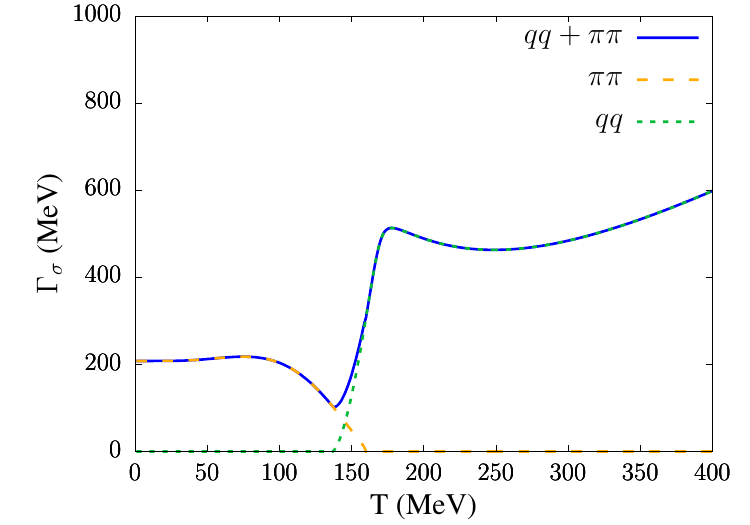}
\caption{The rate for the sigma field to decay into a pair of pions, into a light quark-antiquark pair, and the sum of the rates as functions of temperature for $m_\sigma = 450$ MeV and $T_\pi = 160$ MeV.}
\label{fig:Sigma_rates_Tx160msig450}
\end{figure}

\begin{figure}[h!]
\includegraphics[width=0.7\columnwidth]{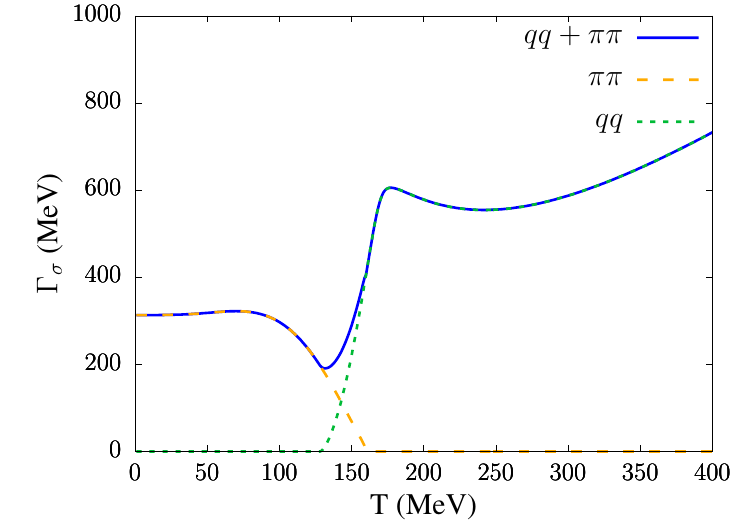}
\caption{The rate for the sigma field to decay into a pair of pions, into a light quark-antiquark pair, and the sum of the rates as functions of temperature for $m_\sigma = 500$ MeV and $T_\pi = 160$ MeV.}
\label{fig:Sigma_rates_Tx160msig500}
\end{figure}

\begin{figure}[t]
\includegraphics[width=0.7\columnwidth]{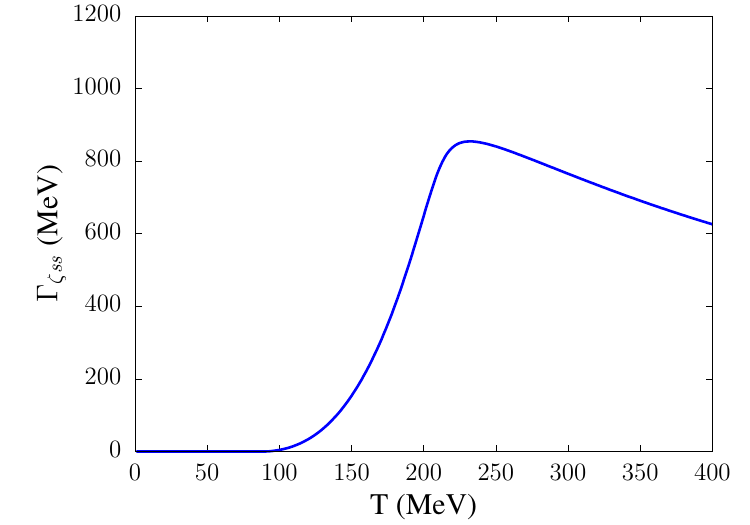}
\caption{The rate for the f$_0$(980) field to decay into a strange quark-antiquark pair as a function of temperature with $g = 3.2$.}
\label{fig:zeta_rate}
\end{figure}

Figures \ref{fig:Sigma_rates_Tx160msig450} and \ref{fig:Sigma_rates_Tx160msig500} display the rate for the sigma field to decay into a pair of pions $\Gamma_{\sigma\pi\pi}$, 
the rate for decay into a light quark--antiquark pair $\Gamma_{\sigma q q}$, and the sum of the rates as functions of temperature $\Gamma_{\sigma}$. 
Figure \ref{fig:zeta_rate} displays the rate for the f$_0$(980) field to decay into a strange quark-antiquark pair as a function of temperature.  The transition is a rapid crossover, not a second-order phase transition, due to $m_l \neq 0$.  Therefore, we have chosen the parameters such that $T_q < T_\pi$.  Note the minimum in $\Gamma_{\sigma}$ near $T = 140$ MeV.

\section{DIC}\label{SecIV}

In this section, we summarize the relationship between the up and down quark condensates and the meson fields. The meson field matrix is $M = {\rm diag} (\sigma_u, \sigma_d, \zeta)$ where $\sigma_u = \sigma \cos\theta$ and $\sigma_d = \sigma \sin\theta$.  In equilibrium $\theta = \pi/4$.  The Lagrangian has the form 
\ba
{\cal L} &=& \thalf {\rm Tr} \left( \partial_\mu M \partial^\mu M^\dagger \right) - U\left( M, M^\dagger \right) \nonumber \\
&=& \thalf \partial_\mu \sigma_u \partial^\mu \sigma_u + \thalf \partial_\mu \sigma_d \partial^\mu \sigma_d
+ \thalf \partial_\mu \zeta \partial^\mu \zeta - U(\sigma_u, \sigma_d, \zeta) \nonumber \\
&=& \thalf \sigma^2 \partial_\mu \theta \partial^\mu \theta + \thalf \partial_\mu \sigma \partial^\mu \sigma
+ \thalf \partial_\mu \zeta \partial^\mu \zeta - U(\sigma, \zeta, \theta)
\ea
The general Lagrange equations of motion for fields $\phi_i$
\be
\partial_\mu \left( \frac{\partial{\cal L}}{\partial (\partial_\mu \phi_i)} \right) - \frac{\partial {\cal L}}{\partial \phi_i} = 0
\ee
lead to
\be\partial_\mu \partial^\mu \sigma_u + \Gamma_u \frac{\partial \sigma_u}{\partial t} = -\frac{\partial U}{\partial \sigma_u}
\ee
and 
\be
\partial_\mu \partial^\mu \sigma_d + \Gamma_d \frac{\partial \sigma_d}{\partial t} = -\frac{\partial U}{\partial \sigma_d}
\ee
where dissipative terms have been added.  This can be checked by considering a spatially uniform system and setting $U = 0$ so that the solution is $\sigma_u(t) = \sigma_0 \exp(-\Gamma_u t) + \bar{\sigma}/\sqrt{2}$, and similarly for the down quark, where the asymptotic value has been assumed. 

Now consider a situation where everything is spatially and temporally uniform except for the field $\theta$.  Multiplying the up equation by $-\sin\theta$, the down equation by $\cos\theta$, and adding them results in
\be
\partial_\mu \partial^\mu \theta + \Gamma_q \frac{\partial \theta}{\partial t} = \frac{1}{\sigma}
\left[ \sin\theta \frac{\partial U}{\partial \sigma_u} - \cos\theta \frac{\partial U}{\partial \sigma_d}\right]
= - \frac{1}{\sigma^2} \frac{\partial U}{\partial \theta}
\label{thetaeq}
\ee
where $\Gamma_u = \Gamma_d \equiv \Gamma_\sigma$ as calculated earlier.  In general, the relevant potential can always be taken in the usefully normalized form $\Delta U(\theta) \equiv U(\theta) - U(\pi/4)$.  For example, if we assume the potential
\ba
U &=&-\frac{1}{2}\mu^2(\sigma_u^2+\sigma_d^2+\zeta^2)+\lambda'(\sigma_u^2+\sigma_d^2+\zeta^2)^2
+ \lambda(\sigma_u^4+\sigma_d^4+\zeta^4) \nonumber \\
&-&2c\sigma_u\sigma_d\zeta -\sqrt{2}c'(m_u\sigma_u+m_d\sigma_d+m_s\zeta).
\ea
where $c(T)$ is defined, following the conventions used in \cite{Kapusta:DIC}, as 
\be
c(T) = c(0)/\left( 1 + 1.2 \pi^{2} \bar{\rho}^{2} T^{2} \right)^{7}
\ee
where $c(0) \equiv 1.732$ GeV and $\bar{\rho} \equiv 0.33$ fm, then
\be
\Delta U(\theta) = \frac{1}{2}\lambda \left[1-\sin^2(2\theta) \right]\sigma^4+c(T) \left[1-\sin(2\theta)\right]\sigma^2\zeta
+f_\pi m_\pi^2 \left[ 1-\frac{\cos\theta+\sin\theta}{\sqrt{2}} \right]\sigma
\ee
and
\be
\frac{\partial \Delta U}{\partial \theta} = - \lambda \sigma^4 \sin 4\theta - 2 c(T) \sigma^2 \zeta \cos 2\theta
+ \frac{f_\pi m_\pi^2 \sigma}{\sqrt{2}} (\sin\theta - \cos\theta)
\ee
where we have used $2 c' m_q = f_\pi m_\pi^2$. See Ref. \cite{Kapusta:DIC} for details. These equations make physical sense.  Note that $\theta = \pi/4$ is a stable solution. If $\theta < \pi/4$ then $\theta$ will increase, while if $\theta > \pi/4$ then $\theta$ will decrease, in both cases relaxing to the equilibrium value due to the dissipative term.

In the Hamiltonian formulation the equations of motion become
\be
p = \sigma^2 \frac{\partial \theta}{\partial t}
\ee
and
\be
\frac{\partial p}{\partial t} + \Gamma_q \, p = \sigma^2 \nabla^2 \theta - \frac{\partial \Delta U}{\partial \theta} + \xi
\label{pieq}
\ee
Here $p$ is the momentum conjugate to $\theta$ and $\xi$ is the noise term.

Without regard to a specific theory it might make sense to write
\be
\Delta U(\theta) = a(T) \left( \theta - \pi/4 \right)^2 + \thalf b(T) \left( \theta - \pi/4 \right)^4
\ee
with $a(T)$ and $b(T)$ left as phenomenological functions.  For the aforementioned potential
\be
\Delta U = \left[ 2 \lambda \sigma^4 + 2 c(T) \sigma^2 \zeta
+ \thalf f_\pi m_\pi^2 \sigma \right] \left( \theta - \pi/4 \right)^2 + \cdot\cdot\cdot
\ee

The noise term must satisfy the fluctuation-dissipation theorem.  Assuming white noise the correlator is
\be
\langle \xi({\bf x}, t) \xi({\bf x}', t') \rangle = D \delta ({\bf x} - {\bf x}') \delta (t - t')
\ee

\section{The strong damping limit}\label{SecV}

If $|{\partial p}/{\partial t}| \ll \Gamma_\sigma |p|$ then we are in the strong damping limit.  In that limit 
\be
\frac{\partial \theta}{\partial t} = \frac{1}{\Gamma_\sigma} \left[ \nabla^2 \theta - 
\frac{1}{\sigma^2} \frac{\partial \Delta U}{\partial \theta } + \xi \right]
\ee
This has exactly the same form as the well-studied Langevin equation.  We can simply translate the variables from textbooks, notably Goldenfeld section 8.3 \cite{Gold}.  See the appendix.  We find that
\be
D = \frac{2\Gamma_\sigma T}{\sigma^2}
\ee
which is dimensionless.  

The noise is generated with the Gaussian probability distribution
\ba
P_\xi(\{\xi({\bf x},t)\}) &=& \frac{1}{N} \exp \left[ - \frac{1}{2D} \int d^3x dt \, \xi^2({\bf x},t)\right] \nonumber \\
N &=& \int [d\xi] \exp \left[ - \frac{1}{2D} \int d^3x dt \, \xi^2({\bf x},t)\right]
\ea
The probability to find $\theta({\bf x})$ at time t is
\be
P_{\theta}(\theta({\bf x}),t) = \langle \delta[\theta({\bf x}) - \theta_\xi({\bf x},t)] \rangle_\xi
\ee
where $\theta_\xi({\bf x},t)$ is the solution to the Langevin equation for a particular realization of 
$\xi({\bf x},t)$.  This probability satisfies the Fokker-Planck equation
\be
\frac{\partial}{\partial t} P_{\theta}(\theta({\bf x}),t) = \frac{1}{\sigma^2 \Gamma_\sigma}
\int d^3x' \, \frac{\delta}{\delta \theta({\bf x}')} \left[ \frac{\delta H_\theta}{\delta \theta({\bf x}')}  P_\theta 
+ T \frac{\delta P_\theta}{\delta \theta({\bf x}')} \right]
\ee
where
\be
H_{\theta} = \int d^3x \left[ \thalf \sigma^2 (\nabla \theta)^2 + \Delta U(\theta) \right] 
\ee
For a uniform system in a volume $V$ the equilibrium solution is
\be
P_\theta(\theta) \propto \exp\left[ - V \Delta U(\theta)/T \right]
\ee
as it should be.

Consider small deviations $\delta\theta$ from the equilibrium value in the form of $\theta = \pi/4 + \delta\theta$.
Then
\be
\frac{\partial \delta\theta}{\partial t} = - \left[ \frac{\delta\theta}{\tau_0} - 
\frac{1}{\Gamma_\sigma} \nabla^2 \delta\theta \right] + \frac{\xi}{\Gamma_\sigma}
\label{first}
\ee
where the long-wavelength relaxation time is
\be
\tau_0 = \frac{\bar{\sigma}^{2} \Gamma_\sigma}{2a(T)}
\ee
After transforming from coordinate $x$ to wave-number $k$ the solution is
\be
\delta\theta \sim {\rm e}^{-t/\tau_k} \, {\rm e}^{ikx}
\ee
where the relaxation time is
\be
\frac{1}{\tau_k} = \frac{1}{\tau_0} + \frac{k^2}{\Gamma_\sigma}
\ee

\section{Inclusion of second order time derivative}\label{SecVI}

Is it legitimate to drop the second-order time derivative?  Keeping it Eq. (\ref{first}) becomes
\be
\frac{\partial^2 \delta\theta}{\partial t^2} + \Gamma_\sigma \frac{\partial \delta\theta}{\partial t} 
= - \frac{\Gamma_\sigma }{\tau_0} \delta\theta + \nabla^2 \delta\theta + \xi
\label{second}
\ee
Dropping the noise term the solution is
\be
\delta\theta \sim {\rm e}^{-\gamma t} \, {\rm e}^{i\omega t} \, {\rm e}^{ikx}
\ee
where
\be 
\gamma = \thalf \Gamma_\sigma 
\ee
and
\be
\omega^2 = k^2 + \frac{\Gamma_\sigma }{\tau_0} - \oneqt \Gamma_\sigma^2
\label{eq:omegaSq}
\ee
The frequency is real, and the system is underdamped when $\tau_0 < 4/\Gamma_\sigma$.  Numerically this is the case for all $T$ under consideration.  

Now the Fokker-Planck equation describes the time evolution of the probability distribution for both the field and its conjugate momentum.
\be
P(p({\bf x}), \theta({\bf x}),t) = \langle \delta\left(p({\bf x}) - p_\xi({\bf x},t)\right)
\, \delta\left(\theta({\bf x}) - \theta_\xi({\bf x},t)\right) \rangle_\xi
\ee
The derivation is similar to the one given in Ref \cite{Gold}.
\ba
\frac{\partial P}{\partial t} &=&
\int d^3x' \left[ \left\langle \frac{\partial p_\xi({\bf x}',t)}{\partial t} \frac{\delta}{\delta p_\xi({\bf x}',t)} 
\delta(p-p_\xi) \delta(\theta - \theta_\xi) \right\rangle \right.\nonumber \\
&+& \left. \left\langle \frac{\partial\theta_\xi({\bf x}',t)}{\partial t} \frac{\delta}{\delta \theta_\xi({\bf x}',t)} 
\delta(p-p_\xi) \delta(\theta - \theta_\xi) \right\rangle
\right] \nonumber  \\
&=& - \int d^3x' \left[ \frac{\delta}{\delta p({\bf x}')} \left\langle \frac{\partial p_\xi({\bf x}',t)}{\partial t}  
\delta(p-p_\xi) \delta(\theta - \theta_\xi) \right\rangle \right.\nonumber \\
&+& \left. \frac{\delta}{\delta \theta({\bf x}')} \left\langle \frac{\partial\theta_\xi({\bf x}',t)}{\partial t}  
\delta(p-p_\xi) \delta(\theta - \theta_\xi) \right\rangle
\right]
\ea
Next the equations of motion for $p_\xi$ and $\theta_\xi$ are used along with the $\delta$--functions to get
\ba
\frac{\partial P}{\partial t} &=&
- \int d^3x' \left\{\frac{\delta}{\delta p({\bf x}')} \left[ \left( -\Gamma p + \sigma^2 \nabla^2 \theta 
- \frac{\partial \Delta U}{\partial \theta} \right) P \right]
+ \frac{\delta }{\delta p({\bf x}')} \left\langle \xi \delta(p-p_\xi) \delta(\theta - \theta_\xi) \right\rangle \right. \nonumber \\
&+& \left. \frac{p({\bf x}')}{\sigma^2}  \frac{\delta P}{\delta \theta({\bf x}')} \right\}
\ea
From here and now on it is understood that $P$ and the $\delta$--functions are to be evaluated at ${\bf x}$ while everything else is evaluated at ${\bf x}'$, in particular the terms within the large parentheses.  Now
\ba
\left\langle \xi({\bf x}',t) \delta(p-p_\xi) \delta(\theta - \theta_\xi) \right\rangle &=&
D \left\langle \frac{\delta}{\delta \xi({\bf x}',t)} \delta(p-p_\xi) \delta(\theta - \theta_\xi) \right\rangle 
\nonumber \\
&=& D \int d^3x'' \left\{ \left\langle \frac{\delta p_\xi({\bf x}'',t)}{\delta \xi({\bf x}',t)}  
\frac{\delta}{\delta p_\xi({\bf x}'',t)} \delta(p-p_\xi) \delta(\theta - \theta_\xi) \right\rangle \right. \nonumber \\
&+& \left. \left\langle \frac{\delta \theta_\xi({\bf x}'',t)}{\delta \xi({\bf x}',t)}  
\frac{\delta}{\delta \theta_\xi({\bf x}'',t)} \delta(p-p_\xi) \delta(\theta - \theta_\xi) \right\rangle \right\}
\ea
where the first equality arises from integrating by parts.  For a given noise the solution to the momentum equation is
\begin{gather}
\nonumber p_\xi({\bf x}'',t) = p_\xi({\bf x}'',0) \hspace{0.1 cm} + \\
\int_0^t dt' \left( -\Gamma_\sigma p_\xi({\bf x}'',t') 
+ \sigma^2 \nabla^2 \theta_\xi ({\bf x}'',t') - \frac{\partial \Delta U}{\partial \theta_\xi}({\bf x}'',t') 
+ \xi({\bf x}'',t') \right)
\end{gather}
One needed derivative is
\be
\nonumber \frac{\delta p_\xi({\bf x}'',t)}{\delta \xi({\bf x}',t'')} = \delta({\bf x}'' - {\bf x}') \theta(t-t'') + \cdot\cdot\cdot
\ee
The additional terms vanish when the limit $t'' \rightarrow t$ is taken, while the step function $\theta(t-t'')$ arises because the noise can only influence the momentum when $t > t''$.  The value of $\theta(0)$ is $\thalf$ as can be verified by taking the limit of a smooth function representing the step function.  Hence
\be
\frac{\delta p_\xi({\bf x}'',t)}{\delta \xi({\bf x}',t)} = \thalf \delta({\bf x}'' - {\bf x}')
\ee
The solution to the other equation of motion is
\be
\theta_\xi({\bf x}'',t) = \theta_\xi({\bf x}'',0) + \int_0^t dt' \tilde{p}_\xi({\bf x}'',t')
+ \int_0^t dt' \int_0^{t'} dt'''\xi({\bf x}'',t''') 
\ee
where 
\be
\tilde{p}_\xi({\bf x}'',t') = p_\xi({\bf x}'',t') - \int_0^t dt' \xi({\bf x}'',t')
\ee
Then 
\be
\frac{\delta \theta_\xi({\bf x}'',t)}{\delta \xi({\bf x}',t'')} = 
\delta({\bf x}'' - {\bf x}') \int_0^t dt' \theta(t' - t'') + \cdot\cdot\cdot =
\delta({\bf x}'' - {\bf x}') (t - t'') + \cdot\cdot\cdot
\ee
This goes to zero as $t'' \rightarrow t$.

Putting everything together results in the Fokker-Planck equation
\ba
\frac{\partial P}{\partial t} &=&
\int d^3x' \left\{ \frac{\delta }{\delta p({\bf x}')} \left[\left( \Gamma p({\bf x}') 
- \sigma^2 \nabla^2 \theta({\bf x}') 
+ \frac{\partial \Delta U}{\partial \theta}({\bf x}') \right) P \right] \right. \nonumber \\
&+& \left. \frac{D}{2} \frac{\delta^2 P}{\delta p^2({\bf x}')} 
- \frac{p({\bf x}')}{\sigma^2}  \frac{\delta P}{\delta \theta({\bf x}')} \right\}
\ea
where again 
\be
P = P(p({\bf x}), \theta({\bf x}),t)
\ee
This ought to be compared to an analogous Fokker-Planck equation in Ref. \cite{Bet} which is similar but which differs in notation and possibly in meaning. 

For a particle of mass $m$ moving in a potential $V(x)$ whose equations of motion are
\ba
m \dot{v} + \gamma v &=& - \frac{d V(x)}{dx} + \xi \nonumber \\
\dot{x} &=& v
\ea
This results in the Langevin equation
\be
\frac{\partial P(x,v,t)}{\partial t} = \frac{\partial}{\partial v} 
\left[\frac{1}{m} \left( \gamma v + \frac{dV(x)}{dx} \right) P + 
\frac{\gamma T}{m^2} \frac{\partial}{\partial v}P \right]
- v \frac{\partial P}{\partial x}
\ee
The equilibrium solution is
\be
P \propto \exp[-(mv^2/2 + V(x))/T]
\ee

The Hamiltonian in the present case is
\be
H = \int d^3x \left[ \frac{p^2}{2\sigma^2} + \thalf \sigma^2 (\nabla \theta)^2 + \Delta U(\theta) \right] 
\ee
and the equilibrium solution should be
\be
P^{\rm eq} \propto \exp(-H/T)
\ee
Using
\be
\frac{\delta P^{\rm eq}}{\delta \theta} = - \frac{1}{T} \frac{\delta H}{\delta \theta} P^{\rm eq} = 
\frac{1}{T} \left( \sigma^2 \nabla^2 \theta 
- \frac{\partial \Delta U}{\partial \theta} \right) P^{\rm eq}
\ee
and
\be
\frac{\delta P^{\rm eq}}{\delta p} = - \frac{1}{T} \frac{\delta H}{\delta p} P^{\rm eq} = 
- \frac{p}{\sigma^2 T} P^{\rm eq}
\ee
we can write
\be
\frac{\partial P^{\rm eq}}{\partial t} = \left( \Gamma - \frac{D}{2 \sigma^2 T} \right) \int d^3x' 
\frac{\delta }{\delta p({\bf x}')} \Big[ p({\bf x}') P \Big]
\ee
because the gradient and potential terms immediately cancel.  Therefore this is a solution if $D = 2 \sigma^2 \Gamma T$.

\section{Harmonic Oscillator}\label{SecVII}

The Fokker-Planck equation derived in the previous section is very challenging to solve for high energy heavy collisions. That is beyond the scope of this paper. In particular, we are mainly interested in the distribution of $\theta$, not in its conjugate momentum. 

Chandrasekhar \cite{Chandrasekhar} solved the problem for the stochastic harmonic oscillator.  With
\be
\frac{d^2x}{dt^2} + 2 \gamma \frac{dx}{dt} + \omega_0^2 x = \xi(t)
\ee
he found the probability distribution in the coordinate $x$ to be
\be
P(x,t) = \left[ \frac{m \omega_0^2}{2\pi T w(t)} \right]^{1/2}
\exp \left[ - \frac{m \omega_0^2}{2T} \frac{(x - \bar{x}(t))^2}{w(t)} \right]
\ee
For the underdamped case
\be
w(t) = 1 - {\rm e}^{- 2\gamma t} \left[ 2 \frac{\gamma^2}{\omega^2} \sin^2 \omega t + 
\frac{\gamma}{\omega} \sin 2\omega t + 1 \right]
\ee
where $\omega = \sqrt{\omega_0^2 - \gamma^2}$.  The function $\bar{x}(t)$ is the solution in the absence of noise with initial position $x_0$ and initial velocity $v_0$ at $t=0$.  Specifically
\be
\bar{x}(t) = {\rm e}^{- \gamma t} \left[ x_0 \cos \omega t + 
\frac{\gamma x_0 + v_0}{\omega} \sin \omega t \right]
\ee
and associated velocity
\be
\bar{v}(t) = {\rm e}^{- \gamma t} \left[ v_0 \cos \omega t - 
\frac{\gamma v_0 + x_0 \omega_0^2}{\omega} \sin \omega t \right]
\ee
Chandrasekhar did not solve a Fokker-Planck equation to get the probability distribution in $x$, instead using another method.  However, knowing the distribution one can determine that it satisfies the equation
\be
\frac{\partial P}{\partial t} = - \bar{v}(t) \frac{\partial P}{\partial x} + 
\frac{1}{2} \frac{T}{m \omega^2} \frac{d w}{dt} \frac{\partial^2 P}{\partial x^2}
\ee

Now let us consider the DIC.  For symmetry, change to the field $\varphi = \theta - \pi/4$.  For $T \ge 180$ MeV, a very good approximation is
\be \label{eq:DeltaUAprx}
\Delta U = \thalf f_\pi m_\pi^2 \bar{\sigma} \, \varphi^2
\ee
because $\bar{\sigma}$ is so small at high temperatures.  Not only is this accurate, but it may be model independent as it avoids fitting any other parameters in the potential $\Delta U$.

\begin{figure}[H]
     \centerline{\includegraphics[width=0.7\textwidth]{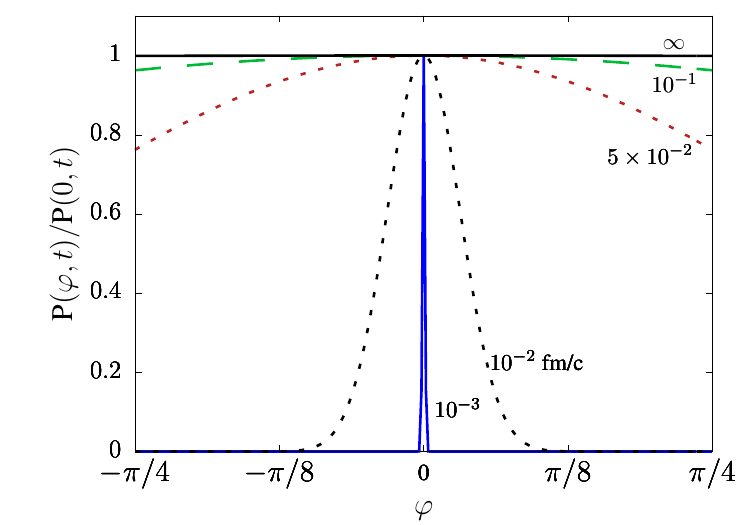}}
     \caption{Temporal evolution of $P(\varphi,t)/P(0,t)$ at V $= 10$ fm$^{3}$, $T = 200$ MeV, $T_{x} = 160$ MeV, $m_{\sigma} = 450$ MeV. Each curve is labeled by the time at which $P(\varphi,t)/P(0,t)$ was plotted.}
     \label{fig:RelativeProb10_200}
\end{figure}

\begin{figure}[H]
     \centerline{\includegraphics[width=0.7\textwidth]{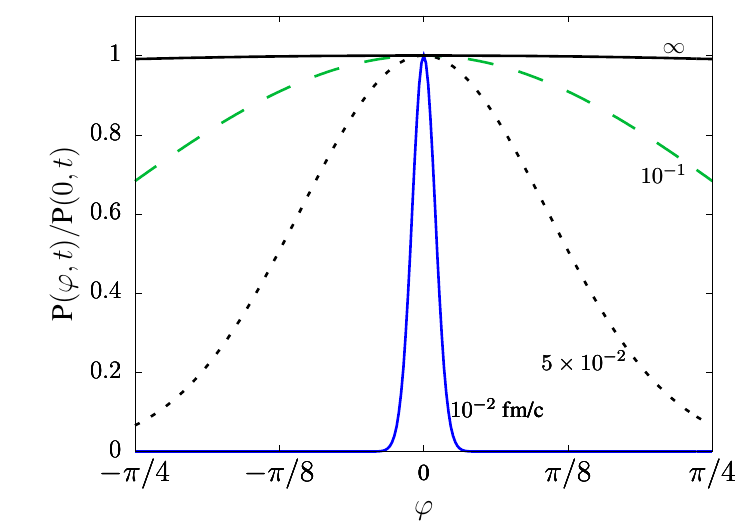}}
     \caption{Temporal evolution of $P(\varphi,t)/P(0,t)$ at V $= 100$ fm$^{3}$, $T = 200$ MeV, $T_{x} = 160$ MeV, $m_{\sigma} = 450$ MeV. Each curve is labeled by the time at which $P(\varphi,t)/P(0,t)$ was plotted.}
     \label{fig:RelativeProb100_200}
\end{figure}

How can we take advantage of the analysis by Chandrasekhar?  One possibility is to consider a volume $V$ within which the field is uniform in space but evolving in time.  The relevant equation is 
\be
\frac{d^2\varphi}{dt^2} + 2 \gamma \frac{d\varphi}{dt} + \omega_0^2 \varphi = \xi(t)
\ee
with
\be
\gamma = \thalf \Gamma_\sigma
\ee
and
\be
\omega_0^2 = \frac{f_\pi m_\pi^2}{\bar{\sigma}}
\ee
The distribution then can be written as
\be
P(\varphi,t) = \left[ \frac{f_\pi m_\pi^2 \bar{\sigma} V}{2\pi T w(t)} \right]^{1/2}
\exp \left[ - \frac{f_\pi m_\pi^2 \bar{\sigma}}{2T} \frac{(\varphi - \bar{\varphi}(t))^2}{w(t)} V\right]
\ee
The temporal evolution of the relative probability
\be
\frac{P(\varphi,t)}{P(0,t)} = \exp \left[ - \frac{f_\pi m_\pi^2 \bar{\sigma}}{2T} \frac{\varphi^2}{w(t)} V\right]
\ee
for $T = 200$ MeV is shown in figures \ref{fig:RelativeProb10_200} ($V = 10$ fm$^3$) and \ref{fig:RelativeProb100_200} ($V = 100$ fm$^3$). The evolution for $T = 190$ MeV and $180$ MeV are very similar.

The distributions quickly reach equilibrium because $\Gamma_\sigma$ is so large.  Even if it was smaller by a factor of 2, 3, or 4, equilibrium would still be reached during the expansion and cooling of the matter produced in high energy heavy ion collisions.  The equilibrium distributions are nearly flat because the potential $\Delta U$ is linear in $\bar{\sigma}$, 
and $\bar{\sigma}$ decreases rapidly for $T > 160$ MeV.

\section{Temporal Evolution}\label{SecVIII}

In an expanding medium the temperature is rapidly changing with time. It is easiest to model the time dependence of the temperature in terms of Bjorken dynamics. For an ideal fluid, the energy density evolves as
\be
\epsilon(\tau) = \epsilon(\tau_0)\left(\frac{\tau_0}{\tau}\right)^{4c_s^2}
\ee
where $c_s$ is the speed of sound. The temperature evolution is
\begin{equation}
    T(\tau) = T(\tau_0)\left(\frac{\tau_0}{\tau}\right)^{c_s^2}
\end{equation}
For an ideal gas equation of state $c_s^2 = 1/3$. 

\begin{figure}[h]
    \centering
    \includegraphics[width=0.7\columnwidth]{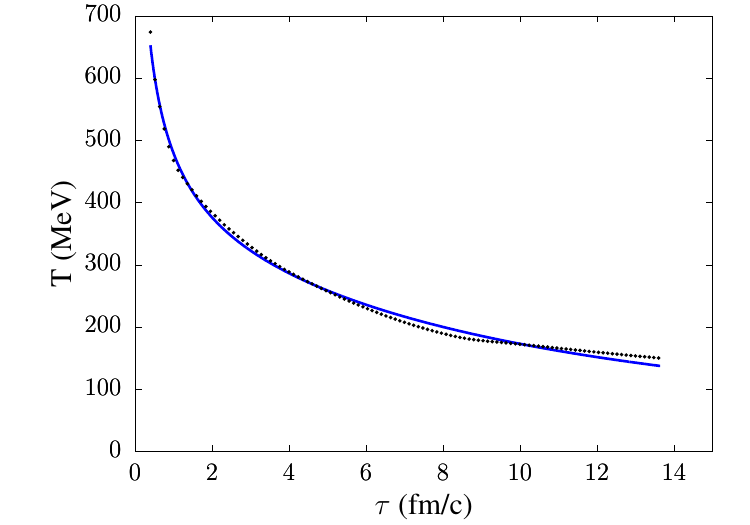}
    \caption{Temperature evolution in the mid-rapidity region in a hydrodynamic simulation of a central Pb-Pb collision at 
    $\sqrt{s_{NN}} = 2.76$ TeV along with the parameterization given in the text.  There is hardly a noticeable difference between them.}  
    \label{fig:TempFitting7_28_alt-2}
\end{figure}

A more realistic parameterization of the temperature evolution is
\be
T^3(\tau) = \frac{X}{(\tau^2 + \tau_{\perp}^2) \tau}
\ee
which takes into account the transition from an initially one-dimensional longitudinal expansion to a three-dimensional expansion which includes the transverse direction.  The above expression was fit to temperature evolution from numerical simulations of Pb-Pb collisions at $\sqrt{s_{NN}} = 2.76$ TeV using the IP-Glasma initial state \cite{Schenke:2012wb,McDonald:2016vlt} and hydrodynamic solver MUSIC \cite{Schenke:2010nt}. The cell with the highest temperature was tracked to extract the temporal evolution and the result was averaged over events from 0-5\% centrality bin. The fit results are shown in Fig. \ref{fig:TempFitting7_28_alt-2}. The parameters are $X = 9.6896 \times 10^{9} \pm 4.58 \times 10^{7}$ and $\tau_{\perp} = 9.3214 \pm 0.0385$ fm/c with $T$ in units of MeV.


\section{Causal Volumes}
\label{SecIX}

The matter initially created in central Pb-Pb collisions at the LHC is at a temperature above 600 MeV.  As it expands and cools the quark condensates will eventually be formed.  Regions of space that are causally disconnected will form condensates independent of each other.  Assuming a Bjorken expansion, Castorina and Satz \cite{CS} calculated the distance beyond which two points could not have communicated via a signal traveling at the speed of light.  The distance in the longitudinal direction is
\be
d = \sqrt{\frac{\tau_2}{\tau_1}} (\tau_2 - \tau_1)
\ee
where the signal is sent at proper time $\tau_1$ and received at time $\tau_2$.  In the transverse direction the corresponding distance is $\tau_2 - \tau_1 \equiv 2r$; this neglects transverse expansion.  We can then estimate the causal volume as that of a rod $V_c = \pi r^2 d$.

Let us take $\tau_1$ as the time at which the light quark condensates begin to form corresponding to the local temperature $T_1$, and $\tau_2$ the time and $T_2$ the temperature at which the probability distribution is no longer flat.  Table \ref{table1} gives some illustrative examples.
\begin{table}
\begin{tabular}{|c|c|c|c|c|c|c|}
\hline
$T_1$(MeV) & $\tau_1$(fm/c) & $\bar{\sigma}_l(T_1)/\sigma(0)$ & $T_2$(MeV) & $\tau_2$(fm/c) & $\bar{\sigma}_l(T_2)/\sigma(0)$ & $V_c$(fm$^3$) \\
\hline
240 & 5.81 & 0.01258 & 180 & 9.44 & 0.06818 & 47.89\\
\hline
220 & 6.82 & 0.01730 & 180 & 9.44 & 0.06818 & 16.61 \\
\hline
200 & 8.01 & 0.02769 & 180 & 9.44 & 0.06818 & 2.49 \\
\hline
240 & 5.81 & 0.01258 & 170 & 10.26 & 0.1802 & 91.97 \\
\hline
220 & 6.82 & 0.01730 & 170 & 10.26 & 0.1802 & 39.21 \\
\hline
200 & 8.01 & 0.02769 & 170 & 10.26 & 0.1802 & 10.12 \\
\hline
240 & 5.81 & 0.01258 & 160 & 11.17 & 0.3932 & 167.70 \\
\hline
220 & 6.82 & 0.01730 & 160 & 11.17 & 0.3932 & 82.74 \\
\hline
200 & 8.01 & 0.02769 & 160 & 11.17 & 0.3932 & 29.27 \\
\hline
\end{tabular}
 \caption{Estimate of causal volumes as described in the text.}
\label{table1}
\end{table}
Clearly there is a large uncertainty in the causal volumes, even more so if one takes $\tau_1$ smaller and the corresponding $T_1$ higher, but they are not inconsistent with phenomenological studies 
\cite{Kapusta:2022ovq,Kapusta:DIC}.

\begin{figure}[H]
    \centering
    \includegraphics[width=0.7\linewidth]{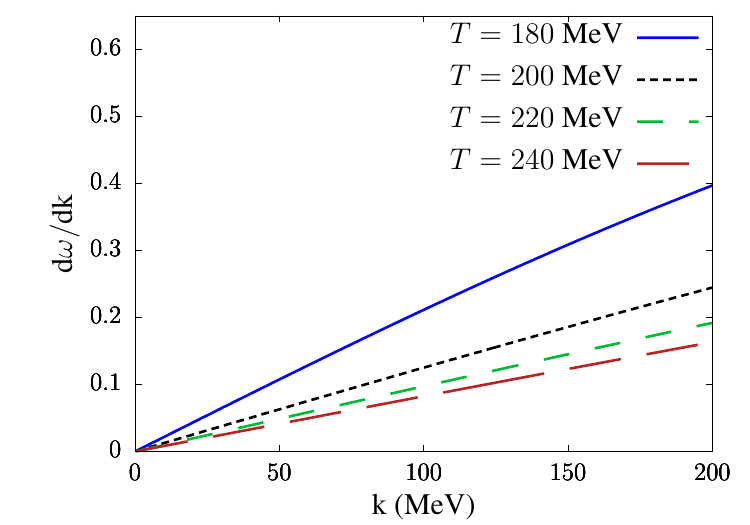}
    \caption{Propagation speed of excitations in $\varphi$}
    \label{fig:vVarphi}
\end{figure}

The signal speed may very well be less than the speed of light, in which case the causal volume would be smaller \cite{CS}.  In the high temperature region $T \geq 180$ MeV, Eq. (\ref{eq:omegaSq}) takes the form
\begin{equation}
    \omega^2 = k^2 + \frac{f_{\pi} m^{2}_{\pi}}{\bar{\sigma}} - \frac{\Gamma_{\sigma}^{2}}{4}
\end{equation}
Making use of this approximation the propagation speed $v = d\omega/dk$ of excitations in $\varphi$ can be computed for temperatures at or above $180$ MeV, as shown in Fig. \ref{fig:vVarphi}.

\section{Summary and conclusions}\label{SecX}

It has previously been shown that the anomalous $\nu_{\rm dyn}$ measurement between the neutral and charged kaons by the ALICE collaboration at $\sqrt{s_{NN}} = 2.76$ TeV cannot be explained by conventional effects. Disoriented isospin condensates have been proposed as a mechanism that can give rise to $\nu_{\rm dyn}$ similar to the experimental data. In this work, we have shown that the DIC is both viable and feasible.

We began with the assumption that the sigma mesons are excited states of the system which decay into a quark and anti-quark pair in QGP and into a pair of pions in hadronic gas. We calculated these decay rates using an effective Lagrangian. We wrote the Langevin and derived the Fokker-Planck equations for the isospin disorientation field $\theta$. We found that for realistic values of the parameters the field is underdamped. These equations are difficult to solve in an expanding system such as produced in heavy ion collisions.  To make progress, we made the assumption that $\theta$ is uniform in space within some volume $V$ and fixed temperature but evolves in time.  The effective potential is essentially quadratic for temperatures $T \ge 180$ MeV.  This allowed us to adapt a solution to the harmonic oscillator by Chandrasekhar.  Starting with an equal mixture of up and down quark condensates, the distribution in $\theta$ relaxed to its equilibrium distribution, which is nearly flat, on a time scale of less than 1 fm/c.  This time scale originates from the width of the sigma meson, and is much less than the expansion time scale.  Hence the conditions for DIC should be well fulfilled at the time of hadronization.

It should be noted that the DIC picture does not require $\theta$ to be uniform across the volume of a domain. The general formalism allows for $\theta$ to vary from point-to-point in position space. As long as the average $\theta$ over a domain volume is non-zero, the phenomenology is recovered. The approximation of a spatially uniform $\theta$ was made to leverage Chandrasekhar's analysis and make quantitative estimations. Within this simplification, the relaxation time is orders of magnitude smaller than the QGP lifetime in these collisions. The DIC would remain viable even if the relaxation times are larger by an order of magnitude. A full dynamic calculation with fluctuating $\theta$ would allow us to test if this holds true.

We argued that the causally connected volumes from the point from when the condensates start emerging to when hadronization happens can range anywhere from about 2 to 200 fm$^3$. This big range reflects the uncertainties associated with the temperatures at which condensates start forming and at which time kaons hadronize. 

In the future, a full numerical space-time evolution of the Langevin or Fokker-Planck equations for the $\theta$ field, coupled with relativistic viscous 3+1 dimensional hydrodynamic equations, can give us more insight into the dynamics of DIC domains.  Lattice QCD should be used to put the underlying theoretical underpinnings on a more rigorous foundation, especially numerically.

\begin{acknowledgments}
We thank S. Pratt and S. Rudaz for feedback on this manuscript. This work was supported by the U.S. DOE Grants No. DE-FG02-87ER40328 (JIK and MS) and  DE-SC-0024347 (MS).
\end{acknowledgments}

\section*{appendix}

The relation between our parameters and those in \cite{Gold} are
\ba
\eta_G &=& \theta - \frac{\pi}{4} \nonumber \\
\zeta_G &=& \frac{\xi}{\Gamma_q} \nonumber \\
\Gamma_G &=& \frac{1}{\sigma^2 \Gamma_q} \nonumber \\
\gamma_G &=& \sigma^2 \nonumber \\
\bar{a}_G &=& a \nonumber \\
b_G &=& b \nonumber \\
D_G &=& \frac{2T}{\sigma^2 \Gamma_q}
\ea
Here the subscript $G$ refers to that book.

\end{document}